\documentclass[11pt]{article}
\usepackage{amsfonts}

\begin{document}
\title{{\bf Extracting molecular Hamiltonian structure from  
time-dependent fluorescence intensity data}}
\author{Constantin Brif and Herschel Rabitz \\
\emph{Department of Chemistry, Princeton University, 
Princeton, New Jersey 08544}}
\date{}
\maketitle

\begin{abstract}
We propose a formalism for extracting molecular Hamiltonian 
structure from inversion of time-dependent fluorescence intensity 
data. The proposed method requires a minimum of \emph{a priori} 
knowledge about the system and allows for extracting a complete 
set of information about the Hamiltonian for a pair of molecular 
electronic surfaces. 
\end{abstract}

\section{Introduction}

A long standing objective is the extraction of molecular 
Hamiltonian information from laboratory data. The traditional 
approaches to this problem attempt to make use of 
time-independent (spectroscopic and scattering) data 
\cite{ibook,HR-inv,ZL95}. 
Another approach aims to use ultrafast temporal data, with 
information on molecular potentials and dipole moments obtained 
for spatial regions sampled by evolving wave packets. Research 
in this direction has been especially intense during 
the last few years \cite{Zew-inv,BK95,Shap,LR95,ZR99a,ZR99b}.
This activity is inspired by recent progress in the technology 
of ultrafast laser pulses \cite{Zew93,SSGMK99}, which makes 
possible observations of molecular dynamics with increasingly 
higher spatial and temporal resolution. 

Due to the difficulty of the Hamiltonian inversion problem, it
is common to assume that some \emph{a priori} knowledge of the
system is available. For example, one technique \cite{Shap} 
proposes to extract time-evolving wave functions and 
excited-state potentials using time-resolved and 
frequency-resolved fluorescence data and knowledge of the 
ground-state potential, the transition frequencies, and the 
transition dipole moment. The inverse tracking method 
\cite{LR95}, proposed for recovering the potential energy and 
dipole moment of a molecular electronic surface by monitoring 
the temporal evolution of wave packets, explicitly assumes 
knowledge of the initial excited wave functions. Clearly, such 
assumptions impair self-consistency and at least partially 
undermine the inversion objectives. Although the desire to 
simplify the inversion algorithm by making \emph{a priori} 
assumptions about what is known and unknown is understandable, 
it has remained an open question about whether these 
assumptions are actually necessary.

This letter addresses the latter point by proposing an inversion 
formalism that makes use of minimal \emph{a priori} knowledge 
about the system. The formalism is designed to operate between 
two electronic surfaces, with electronic and vibrational 
transitions driven by two fast laser pulses, which allows for 
extracting the potential energies and dipole moments for both 
surfaces as well as the electronic transition dipole moment. 
The extraction is based on the inversion of the time-dependent 
fluorescence intensity data obtained from the detection of 
spontaneous emission in transitions between the electronic 
surfaces. The proposed formalism lays the ground work for 
extracting a complete set of information about a pair of 
electronic surfaces in a closed way, with a minimum of 
\emph{a priori} assumptions about the molecular Hamiltonian. 
This letter presents the conceptual foundation of this novel 
approach, and a detailed numerical algorithm with simulations 
will be presented elsewhere.

\section{The physical picture}

Consider the ground and excited electronic molecular surfaces 
with potential energies $V_g (x)$ and $V_e (x)$ and dipole 
moments $\mu_g (x)$ and $\mu_e (x)$, respectively. The dipole 
moment for the electronic transition between the two surfaces 
is $M(x)$. For the sake of conceptual clarity, we consider a
one-dimensional problem; the generalization for the 
multidimensional case is straightforward.

The setup includes two time-dependent locked laser fields: 
$\epsilon_0 (t)$ drives transitions between the two 
electronic surfaces (the carrier frequency of this laser 
will be typically in the visible or ultraviolet part of the 
spectrum), and $\epsilon_1 (t)$ drives transitions between 
vibrational levels within each of the two surfaces (the carrier 
frequency of this laser will be typically in the infrared). 
The role of the driving fields is to excite the molecular wave 
packet and guide its motion on the surfaces. It is physically 
reasonable that the potentials and dipole moments may be 
reliably extracted only in the region sampled by the evolving 
wave packet. We assume that interactions with other electronic
surfaces and incoherent processes (e.g., thermal excitation 
and collisional relaxation) are negligible. 
Ultrafast laser technology has made great advances recently, 
but preparation of the infrared pulse of a desired shape is 
still a challenging technical problem.
We will consider the general situation, with two locked driving
fields and five unknown functions (two potentials and three
dipole moments), but taking $\epsilon_1 =0$ the problem is 
easily reduced to a simpler one, with only one driving field 
$\epsilon_0$ and three unknown functions ($V_g$,  $V_e$ and 
$M$). 

The Hamiltonian of the system in the Born-Oppenheimer, 
electric-dipole and rotating-wave approximations takes the 
form:
\begin{equation}
H = H_g (x,p,t) \sigma_{g g} + H_e (x,p,t) \sigma_{e e} - 
M(x) \epsilon_0 (t) (\sigma_{e g} + \sigma_{g e}) ,
\end{equation}
where $\sigma_{i j} = | i \rangle \langle j |$ 
(with $i, j = g, e$) are transition-projection operators for
the electronic states $| g \rangle$ and $| e \rangle$. Here,
$x$ and $p$ are the canonical position and momentum for the
vibrational degree of freedom, $H_g$ and $H_e$ are the 
vibrational Hamiltonians in the ground and excited electronic
states,
\begin{equation}
H_i (x,p,t) = T + V_i (x) - \mu_i (x) \epsilon_1 (t) ,
\hspace{8mm} i = e,g ,
\end{equation}
and $T = p^2 /2m$ is the kinetic energy of the vibrational 
motion.

We assume that the initial state of the system is
$|\Psi (0) \rangle = | u_0 \rangle | g \rangle$,
where $u_0 (x)$ is the vibrational ground state localized in
the known harmonic part of the potential $V_g (x)$.
The state of the system at any time $t$ will be of the form
\begin{equation}
\label{vfform}
|\Psi (t) \rangle = |u(t)\rangle |g\rangle + 
|v(t)\rangle |e\rangle ,
\end{equation}
with the normalization condition
$\int d x \left( | u(x,t) |^2 + | v(x,t) |^2 \right) = 1$.
The Schr\"{o}dinger equation,
$ i \hbar \partial_t |\Psi (t) \rangle =  H 
|\Psi (t) \rangle$,
then takes the form
\begin{eqnarray}
\label{SEu}
& & i \hbar \partial_t u(x,t) = 
- \frac{ \hbar^2 }{2 m} \partial_x^2 u(x,t) + 
[V_g (x) - \mu_g (x) \epsilon_1 (t)] u(x,t) 
- M(x) \epsilon_0 (t) v(x,t) , \\
\label{SEv}
& & i \hbar \partial_t v(x,t) = 
- \frac{ \hbar^2 }{2 m} \partial_x^2 v(x,t) + 
[V_e (x) - \mu_e (x) \epsilon_1 (t)] v(x,t) 
- M(x) \epsilon_0 (t) u(x,t) ,
\end{eqnarray}
with the initial conditions $u(x,0) = u_0 (x)$, $v(x,0) = 0$.

The radiation emitted spontaneously by the molecule via
transitions between the excited and ground electronic surfaces
contains information about the wave packet. This 
fact has been used to reconstruct unknown vibrational wave 
packets in the method of emission tomography \cite{WW98}. 
Our aim is different: we assume that the initial state of the 
system is known and want to extract the unknown potentials 
($V_g$ and $V_e$) and dipole moments ($\mu_g$, $\mu_e$, and $M$) 
from information contained in the time-dependent fluorescence. 
We choose the time-dependent intensity of the emitted radiation,
$I(t)$, as the observable. This intensity is 
$I(t) = E^{(+)} (t) E^{(-)} (t)$, where $E^{(+)}(t)$ is the 
negative-frequency part of the electric field operator of the
emitted radiation. $E^{(+)}$ is proportional to 
$M \sigma_{e g}$, so the measured quantity is
\begin{equation}
\langle I(t) \rangle = \kappa \langle \Psi (t) | 
M^2 \sigma_{e e} | \Psi (t) \rangle .
\end{equation}
where $\kappa$ is a proportionality constant. 

\section{Extraction of the Hamiltonian structure}

The physical picture above leads to the the following 
mathematical problem: extract the potentials and dipole moments 
from the measured intensity $\langle I(t) \rangle$, assuming 
that the initial state and the two driving fields are known
(note that a number of advanced experimental techniques have been 
recently developed for characterization of ultrashort optical 
pulses \cite{frog,trog,spider}). 

We start from the Heisenberg equation of motion,
$ i \hbar d I/ d t = [I, H]$, to obtain
\begin{equation}
\label{1Heq}
\frac{ i \hbar }{\kappa} \frac{ d \langle I \rangle}{ d t } = 
\langle \Psi(t)| [M^2, T] \sigma_{e e} - \epsilon_0 (t) 
M^3 (\sigma_{e g} - \sigma_{g e}) | \Psi (t) \rangle .
\end{equation}
Using form (\ref{vfform}) of the wave function, we rewrite 
(\ref{1Heq}) as an integral equation for $M(x)$:
\begin{equation}
\label{1ieq}
\int d x [ M^2 (x) F(x,t) + M^3 (x) G(x,t) ] = 
\frac{\hbar}{2 \kappa} \frac{d \langle I \rangle}{ d t } ,
\end{equation}
where
\begin{equation}
F(x,t) = \frac{ \hbar^2 }{2 m} {\mathrm{Im}} 
[ v(x,t) \partial_x^2 v^{\ast}(x,t) ] , \hspace{8mm}
G(x,t) = \epsilon_0 (t) {\mathrm{Im}}
[ u^{\ast}(x,t) v(x,t) ] .
\end{equation}

In order to obtain equations for the other unknown 
functions (two potentials and two dipole moments), we consider 
the second time derivative of $\langle I(t) \rangle$.
Then, using (\ref{vfform}), we derive the following integral 
equation:
\begin{eqnarray}
\frac{ \hbar^2 }{\kappa} \frac{ d^2 \langle I \rangle}{ d t^2 } 
+ T_M (t) 
& = & \int d x [V_e (x) - V_g (x) - \epsilon_1 (t) \mu_e (x)
+ \epsilon_1 (t) \mu_g (x)] S_M (x,t) \nonumber \\
& & + \int d x [V_e (x) - \epsilon_1 (t) \mu_e (x)] R_M (x,t) ,
\label{2ieq}
\end{eqnarray}
where
\begin{eqnarray}
& & R_M (x,t) = \frac{ \hbar^2 }{m} {\mathrm{Re}} \left\{ 
v^{\ast} (x,t) [\partial_x^2 , M^2 (x) ] v(x,t) \right\} , \\
& & S_M (x,t) = - 2 \epsilon_0 (t) M^3 (x) {\mathrm{Re}}
\left[ u^{\ast} (x,t) v^{\ast} (x,t) \right] , 
\end{eqnarray}
\begin{eqnarray}
T_M (t) & = & \frac{ \hbar^4 }{ 4 m^2 } 
\int d x\, v^{\ast} (x,t) [M^2 (x) \partial_x^4 - 
2 \partial_x^2 M^2 (x) \partial_x^2 + 
\partial_x^4 M^2 (x)] v(x,t) \nonumber \\
& & + \frac{ \hbar^2 }{m} \epsilon_0 (t) {\mathrm{Re}} \left\{ 
\int d x v^{\ast} (x,t) [M^3 (x) \partial_x^2 + 
M^2 (x) \partial_x^2 M (x) - 2 \partial_x^2 M^3 (x) ] 
u(x,t) \right\} \nonumber \\
& & + 2 \epsilon_0^2 (t) \int d x M^4 (x) \left( | v(x,t) |^2 
- | u(x,t) |^2 \right) .
\end{eqnarray}
It is convenient to formally enumerate the unknown functions,
\begin{equation}
f_1 (x) = V_g (x) , \hspace{6mm}
f_2 (x) = V_e (x) , \hspace{6mm}
f_3 (x) = d_0 \mu_g (x) , \hspace{6mm}
f_4 (x) = d_0 \mu_e (x) ,
\end{equation}
where $d_0 = 1$ V/m, so all the functions $f_r (x)$ have the 
dimension of energy. Then the integral equation (\ref{2ieq}) 
takes the form
\begin{equation}
\label{2ieq-m}
\int d x \sum_{r=1}^{4} K_r (x,t) f_r (x) = g(t) ,
\end{equation}
where
\begin{eqnarray}
& & K_1 (x,t) = - S_M (x,t) , \hspace{8mm}
K_2 (x,t) = R_M (x,t) + S_M (x,t) , \\
& & K_3 (x,t) = - \tilde{\epsilon}_1 (t) K_1 (x,t) , \hspace{8mm}
K_4 (x,t) = - \tilde{\epsilon}_1 (t) K_2 (x,t) , \\
& & g(t) = \frac{ \hbar^2 }{\kappa} 
\frac{ d^2 \langle I \rangle}{ d t^2 } + T_M (t) ,
\end{eqnarray}
and $\tilde{\epsilon}_1 = d_0^{-1} \epsilon_1$ is the scaled
(dimensionless) field. 

It is important to emphasize that in fact equations~(\ref{1ieq})
and (\ref{2ieq-m}) represent an infinite number (or, in practice, 
a large number) of equations corresponding to different times. 
We will use this fact in the regularization procedure below.
Of course, equation~(\ref{2ieq-m}) is nonlinear because the wave 
function depends on the potentials and dipole moments. Similarly, 
a solution $M$ of (\ref{1ieq}) depends on the wave function and 
thereby depends on other unknown functions. Consequently, the 
problem at hand, including the integral equations and the
Schr\"{o}dinger equation, is highly nonlinear. 
More importantly, the solution for such a system of integral 
equations is generally not unique and the problem is ill-posed 
(i.e., the solution is not stable against small changes of the 
data). These characteristics are common to virtually all
inverse problems and arise because the data used for the 
inversion are inevitably incomplete. Consequently, we need to 
regularize the problem by imposing physically motivated 
constraints on the unknown functions. For example, we may use 
the fact that physically acceptable potentials and dipoles 
should be smooth functions and tend to zero asymptotically as 
$x \rightarrow \infty$ (in the case of the dipole, the atoms
are assumed to separate as neutrals). By taking into account 
this information, some constraints are imposed on the solutions, 
singling out the functions with desirable physical properties. 
This regularization procedure will stabilize the solution.

The regularized solution of equation~(\ref{2ieq-m}) is achieved 
by minimizing the functional
\begin{equation}
\label{Jfunc}
{\mathcal{J}} = \int_{0}^{t} d t' \left[ \int d x 
\sum_{r=1}^{4} K_r (x,t') f_r (x) - g(t') \right]^2 + 
\sum_{r=1}^{4} \alpha_r \int d x f_r^2 (x) .
\end{equation}
Here, $\alpha_r$ are standard regularization parameters which 
denote the tradeoff between reproducing the laboratory data and 
obtaining the solution with smooth and regular functions. 
The time integration in (\ref{Jfunc}) has a simple physical 
meaning: the measured intensity brings in information about the 
potentials and dipoles at each instance of time and we want to 
use all the laboratory information which has been accumulated 
during the period from time zero until $t$. 
The choice of the functional (\ref{Jfunc}) is not unique, and 
other forms of regularization may be considered as well.

Taking the variation of the functional ${\mathcal{J}}$ with 
respect to the unknown functions $f_r (x)$ involves a subtlety
related to the nonlinearity of the problem: the kernels $K_r$
and the free term $g$ depend on the wave function and on
$M(x)$ and thereby depend on $f_r (x)$. The practical
(numerical) solution of any nonlinear problem includes some
kind of linearization. Here, the point at which we make the
linearization is determined in the regularization procedure.
We choose to take the variation of the functional 
${\mathcal{J}}$ in equation~(\ref{Jfunc}) only with respect to 
the \emph{explicit} dependence on $f_r (x)$. 
Then we obtain the set of regularized equations:
\begin{equation}
\label{req-f}
\int d x' \sum_{r=1}^{4} {\mathfrak{K}}_{p r} (x,x',t) f_r (x')
+ \alpha_p f_p (x) = {\mathfrak{g}}_p (x,t) ,
\end{equation}
where
\begin{eqnarray}
& & {\mathfrak{K}}_{p r} (x,x',t) = \int_0^t d t' 
K_p (x,t') K_r (x',t') , \\
& & {\mathfrak{g}}_p (x,t) = \int_0^t d t'
K_p (x,t') g(t') .
\end{eqnarray}
With $p,r = 1,2,3,4$, we have the system of four integral 
equations with four unknown functions (two potentials and two
moments).

Now we want to regularize equation~(\ref{1ieq}) for the electronic
transition dipole $M$. This equation is highly nonlinear: in 
addition to the dependence on $M$ in the wave function, it also
involves second and third powers of $M$. Once again, we may
choose at which point to make the linearization. We prefer to 
linearize at an early stage, in order to obtain an equation of 
a simple form. Thus we define
\begin{equation}
F_M (x,t) = M^2 (x) F(x,t) , \hspace{8mm}
G_M (x,t) = M^2 (x) G(x,t) ,
\end{equation}
and write the functional
\begin{equation}
\label{JMfunc}
{\mathcal{J}}_M = \int_{0}^{t} d t' \left[ 
\int d x M(x) G_M (x,t') + \int d x F_M (x,t') - 
g_M (t') \right]^2 + \alpha_M \int d x M^2 (x) .
\end{equation}
Here, $g_M (t))$ is the right-hand side of equation~(\ref{1ieq}).
The regularized solution of equation~(\ref{1ieq}) is achieved by
minimizing this functional. And we choose the linearization 
procedure by taking the variation of ${\mathcal{J}}_M$ in 
equation~(\ref{JMfunc}) only with respect to the explicit 
dependence on $M$ (that is, we treat $F_M$ and $G_M$ as
independent of $M$). Then we obtain 
\begin{equation}
\label{req-M}
\int d x'  {\mathfrak{K}}_M (x,x',t) M (x')
+ \alpha_M M (x) = {\mathfrak{g}}_M (x,t) ,
\end{equation}
where
\begin{eqnarray}
& & {\mathfrak{K}}_M (x,x',t) = \int_0^t d t' 
G_M (x,t') G_M (x',t') , \\
& & {\mathfrak{g}}_M (x,t) = \int_0^t d t'
G_M (x,t') \left[ g_M (t') - \int d x' F_M (x',t') \right] .
\end{eqnarray}
Finally, the integral equations (\ref{req-f}) and (\ref{req-M})
for the potentials and dipole moments and the 
Schr\"{o}dinger equations (\ref{SEu}) and (\ref{SEv}) for the
components of the wave function form the full set of coupled 
equations for the unknown functions, with $\langle I(t) \rangle$,
$\epsilon_0 (t)$, and $\epsilon_1 (t)$ as input data.

We conclude the presentation of the formalism with a schematic 
outline of the inversion algorithm which will be numerically 
implemented in a forthcoming work. First, the algorithm will 
start with trial functions for the potentials and dipoles to 
propagate the wave function from $t=0$ to $t = \Delta t$.
Using trial functions at the first step is not an excessive
demand for two reasons: (i) For a sufficiently small
time increment $\Delta t$, the evolution of the wave function
is mainly affected by the values of the potentials and dipoles 
in the region where $u_0 (x)$ is localized, i.e., in the harmonic 
region of the ground potential surface; such information is 
usually known with reasonable accuracy. (ii) As more data becomes 
available, the initial trial functions will be replaced by those 
which match the measured fluorescence intensity. 
The second step will use the measured fluorescence intensity, the 
wave function components $u(\Delta t)$ and $v(\Delta t)$, and the 
initial trial functions for solving equations~(\ref{req-M}) and 
(\ref{req-f}) to obtain the next evaluation of the potentials and 
dipoles. These functions will be once again substituted into the 
Schr\"{o}dinger equation to propagate the wave packet from 
$t = \Delta t$ to $t = 2 \Delta t$. The procedure will be repeated 
many times with new laboratory data incorporated at each time step. 
The recorded fluorescence intensity, $\langle I(t) \rangle$, 
contains information about the potentials and dipoles in the 
region where the wave packet is localized at moment $t$ as well 
as where it was prior to that time. The sequential marching 
forward in time over the data track acts to refine the potentials 
and dipoles at each time step. 

\section{Discussion}

This letter sets forth the formalism of a novel comprehensive 
approach to the inversion of molecular dynamics from 
time-dependent laboratory data. 
One of the main features of the proposed inversion method is 
initiation in the well-known ground state and use of external 
driving fields to excite the wave packet and guide its motion on 
the ground and excited potential surfaces. 
Different driving fields will induce different dynamics and may be
more or less helpful for the inversion procedure. Consequently,
we are left with the attractive prospect of choosing the driving 
fields to be optimally suited for assisting the extraction of 
unknown potentials and dipoles from laboratory data. This choice
may be facilitated by a closed learning loop \cite{JR92} in the
laboratory, starting with a number of different trial 
fields. According to the inversion objectives, a learning 
algorithm will determine the best candidates and direct the fields 
to shapes which are best suited to produce these objectives.
Natural objectives are to maximize the spatial region where the 
potentials and dipoles are reliably extracted. Physical intuition 
suggests that one may learn more about the Hamiltonian at a 
specific spatial point if the wave packet is not spread over the 
whole potential surface but is essentially localized in a narrow 
region around this point. Consequently, the driving fields best 
suited for the inversion will control the dispersion of the wave 
packet and guide its motion in a desired large spatial region.
A numerical simulation of the algorithm, including its
closed-loop learning features, will be the next step towards its
laboratory implementation.

\section*{Acknowledgments}

This work was supported by the U.S. Department of Defense
and the National Science Foundation.

%\newpage
%\section*{References}

\end{document}